\documentclass[aps,pre,reprint,superscriptaddress,longbibliography]{revtex4-1}
\usepackage{amssymb}
\usepackage{amsmath}
\usepackage{cases}
\usepackage{graphicx,color}
\usepackage{hyperref}  % or \usepackage{url}
\definecolor{r}{rgb}{1,0,0}   
\definecolor{g}{rgb}{0,1,0}   
\definecolor{b}{rgb}{0,0,1}

\begin{document}

%\preprint{}

\title{Effective exponents for the diffusive coarsening of wet foams and analogous materials}


%\author{Douglas J.~Durian}
%\thanks{}
%\affiliation{}

\author{Douglas J. Durian}
\email{djdurian@physics.upenn.edu}
\affiliation{Department of Physics and Astronomy, University of Pennsylvania, Philadelphia, PA 19104, USA}
%\affiliation{Center for Computational Biology, Flatiron Institute, Simons Foundation, New York, NY 10010, USA}
\affiliation{Department of Mechanical Engineering and Applied Mechanics, University of Pennsylvania, Philadelphia, PA 19104}


\date{\today}

\begin{abstract}
It is a long-standing puzzle why experimentally measured exponents for power law growth of the average bubble radius cross over from 1/2 in the dry limit to 1/3 near the jamming transition, rather than in the very wet limit of dilute bubbles. Here, this is explained by calculation of the diffuse gas exchange rate between nearly kissing spherical bubbles. The result is only logarithmically faster than a 1/3 power law, a difference too small to be seen in existing data.  The same approach is generalized to jammed foams, where bubbles are modeled as truncated spheres with circular soap films of average radii $fR$ where $R$ is the average bubble radius. The result is a non-power law growth law that gives an effective exponent that varies from 1/2 toward 1/3 as $f$ decreases to 0 at the jamming transition. This exponent prediction is compared with existing data sets in terms of a system-specific connection of $f$ to liquid content.
\end{abstract}

%\pacs{}
%\keywords{}

\maketitle

%\section{}
%\subsection{}
%\subsubsection{}

% If in two-column mode, this environment will change to single-column
% format so that long equations can be displayed. Use
%\begin{widetext}
%\end{widetext}

%-----------------------------------------------
%-------------------------------------------------
\section{Introduction}

Foams are nonequilibrium dispersions of gas bubbles that can evolve by three different mechanisms, all affecting applications~\cite{WeaireHutzlerBook, Cantat2013, Langevin2020}. Even in the absence of gravitational drainage and abrupt bubble-bubble coalescence via film rupture, there is an inevitable diffusion of gas from higher pressure bubbles into neighboring lower pressure bubbles that tend to be larger due to Laplace's Law.  This is driven by surface tension, and serves to reduce the total interfacial area as the average bubble size grows and the texture of the foam coarsens. Similar behavior happens in other phase-separating systems like binary liquids, emulsions, and metal alloys.  The usual practice is to assume that the bubble / droplet / grain domain size distribution evolves into a statistically self-similar scaling state where domain shapes and dimensionless moment ratios are time-independent and the average domain radius asymptotes to a power law $R\sim t^\beta$ at long times~\cite{Mullins1986}. For early times, data can be modeled as
\begin{equation}
    R(t) \approx [\mathcal{D}t+(R_o-\delta)^{1/\beta}]^\beta + \delta
\label{approach}
\end{equation}
where $\mathcal{D}$ is an overall rate constant, $R_o$ is the initial average bubble radius at $t=0$ and $\delta$ is a length set by the difference in width of the initial bubble size distribution from that in the statistically self-similar scaling state~\cite{Chieco2023, delta}. The sign of $\delta$ is positive if the initial state is too polydisperse, and negative if it is too monodisperse.

For highly compressed grains and very dry foams, where domains appear polyhedral and ideally fill space, the coarsening exponent is exactly $\beta=1/2$~\cite{VonNeumann, Mullins56, Mullins1986, Stavans93, MacPhersonSrolovitz2007, Stevenson2010, GlazierGraner2010}.  For very dilute spherical domains where the interstitial medium nearly fills space, the coarsening exponent is exactly $\beta=1/3$~\cite{Lifshitz1961, Wagner1961, VoorheesJSP85, Mullins1986, Taylor1998}.  In terms of Lifshitz-Slyozov Wagner theory \cite{Lifshitz1961, Wagner1961}, the latter is based on a long-tailed $1/r$ radially-symmetric decay of the concentration of dissolved gas away from each bubble (in three dimensions). Therefore, the average bubble separation ought to be much greater than the length of this tail for the $\beta=1/3$ dilute bubble limit to hold. Similarly, in terms of Mullins' scaling arguments \cite{Mullins1986}, $\beta=1/3$ should not hold whenever the soap film thickness $\ell$ plays a role -- as it does at and even below jamming where it sets the surface-surface separation between a multitude of ``contacting" neighbors.

Actual materials lie between the very wet and very dry extremes.  It's an ongoing challenge to account for trends in behavior as a function of wetness, \textit{i.e.} versus liquid fraction $\varepsilon$ or gas bubble packing fraction $\phi=1-\varepsilon$. Many works have addressed how the rate constants $\mathcal{D}$ change for systems close enough to the dilute bubble \cite{VoorheesJSP85, GlicksmanAM12} or the dry foam \cite{HutzlerWeaire00, Stone01, Feitosa08, Graner2012, PasquetSM2023, Morgan2026} limits that the respective exponents can be safely taken as 1/3 or 1/2 independent of liquid content. The latter employ a ``border blocking" approach~\cite{HutzlerWeaire00, BoltonWeaire91} wherein liquid-inflated Plateau borders are assumed to totally block all gas diffusion except across the soap films.  However, ``border-crossing" calculations show that significantly more gas is transported outside the films than would be expected based on average border thickness~\cite{Schimming2017}.

Another approach, in addition to considering rate constants $\mathcal{D}$ with fixed exponents, is to analyze coarsening data empirically in terms of fitted growth exponents that can vary continuously with liquid content~\cite{DurianWeitzPine91b, Graner2012, Isert2013, Graner2015, PuistoPRE2018, GlicksmanAM22}.  Contrary to expectation, $\beta$ was measured for foams of increasing wetness to approach 1/3 near the jamming transition near $\varepsilon =36\%$ liquid, where the bubbles are close-packed spheres \cite{Isert2013}, rather than in the dilute-bubble limit $\varepsilon\rightarrow 1$.  The same confounding behavior was also seen for a sequence of foams recently measured aboard the International Space Station (ISS)~\cite{CRAS2023} but at a different liquid fraction.  This puzzling crossover at the jamming transition remains unexplained, and has broad implications for our understanding coarsening behavior in systems other than just aqueous foams.

Here, the puzzle of $\beta\rightarrow 1/3$ near the jamming transition is resolved using the ``border-crossing" approach to account for the diffuse gas current between neighboring bubbles outside the films~\cite{Schimming2017}. Since the bubble-bubble contact geometry is too complex for analytically solving the diffusion equation, the crux is to make a stratified approximation of the dissolved gas concentration field.  This was shown to give accurate diffuse gas currents by comparison with numerical solutions of the diffusion equation~\cite{Schimming2017} in relevant model geometries. Below, this approach is invoked for nearly-kissing spheres, \textit{i.e.} for foams at the jamming transition. Then it is generalized to spherical bubbles truncated by circular soap films, \textit{i.e.} for jammed foams with variable degrees of wetness as parameterized by the average radius $\rho_f=fR$ of the soap films and how the dimensionless parameter $f$ varies with liquid content. The predicted average bubble growth is not a pure power law, but nonetheless gives an effective exponent that agrees well with data and can be understood physically in terms of a key geometrical feature of the bubble-scale microstructure that changes with liquid content.  Namely, because of border-crossing physics, $\beta$ decreases from 1/2 toward 1/3 at jamming simply because that is where $f$ vanishes.  Interestingly, experimentally observed differences in the crossover of $\beta$ can be reconciled in terms of differences in the form of $f$ versus liquid content that arise from system-specific features such as degree of bubble-bubble adhesion.

%-------------------------------------------------
%-------------------------------------------------
\section{Coarsening at jamming}

%=====================
\begin{figure}[t!]
\includegraphics[width=3.20in]{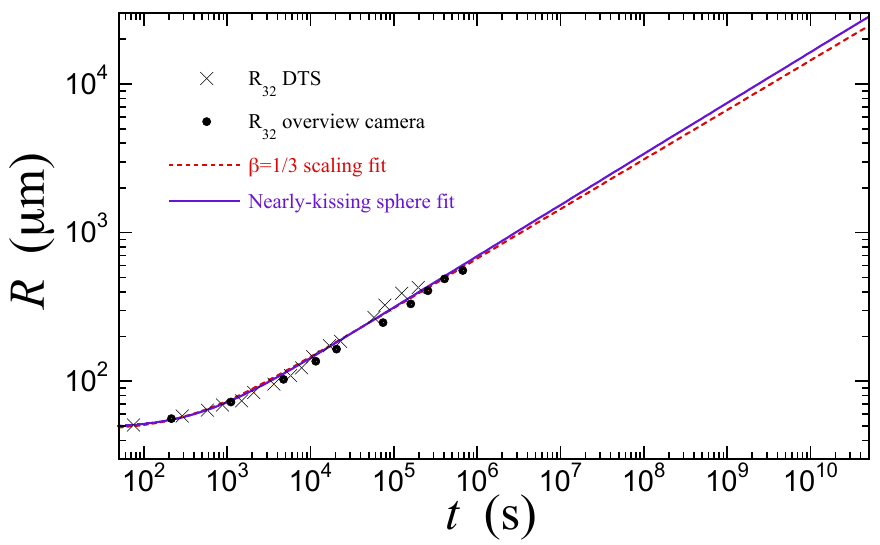}
\caption{Average bubble radius versus time data for the $\varepsilon=0.5$ foam in Fig.~10 of Ref.~\cite{CRAS2023}, along with fits to Eq.~(\ref{approach}) (dotted) with $\beta=1/3$ and $\delta=0$ and to Eq.~(\ref{kissing}) with $\ell=30$~nm (solid). The quality of fit for the latter is unchanged if $\ell$ is varied by a factor of five. The $x$ and $y$ axes are extended to emphasize the near-impossibility of obtaining data that could distinguish the two forms.}
\label{fig_kissing}
\end{figure}
%=====================

At the jamming transition in three dimensions, gas bubbles are nearly-kissing spheres separated by no less than the equilibrium film thickness, $\ell$. For this geometry, numerical solution of the total gas current $I_s$ between neighbors of radii $r_1$ and $r_2$ is well-approximated by the border-crossing prediction of Eq.~(28) computed in Ref.~\cite{Schimming2017} (and repeated for $r_1\approx r_2$ below):
\begin{eqnarray}
\frac{I_s}{D\Delta\phi} &\approx& \pi \frac{r_1r_2}{\langle r\rangle}\ln\left(\frac{r_1r_2}{\langle r\rangle\ell}\right)\label{Is} \\
\Delta\phi &=& \left( \frac{2\gamma}{r_1}-\frac{2\gamma}{r_2}\right) H \label{Dphi}
\end{eqnarray}
where $\langle r\rangle=(r_1+r_2)/2$, $D$ is gas diffusivity, $\gamma$ is the liquid-gas surface tension, $H$ is Henry's Law constant, and $\Delta\phi$ is the difference in gas concentration just outside the two gas bubbles due to their Laplace pressure difference.  Due to this gas current, one bubble shrinks and the other grows with opposite rates $dV/dt=\pm I_s$ where $V=(4/3)\pi r_i^3$.  For a full theory of coarsening, this would be used to determine the evolution of a distribution of bubbles in a packing.  An actual packing would also have a range of neighboring bubbles whose surfaces are separated by distances $l>\ell$.  For such pairs, $\ell$ in Eq.~(\ref{Is}) would be replaced by $l$.

Here, interest is in the average growth rate presuming a statistically self-similar scaling state where all dimensionless bubble size moment ratios are constant. Then, the average bubble volume grows as $dR^3/dt \propto I_s$ where $R$ is an average bubble radius while $r_1$ and $r_2$ are characteristic of different radii from the full distribution. Taking $r_1\propto R$ and $r_2\propto R$ in Eqs.~(\ref{Is}-\ref{Dphi}) gives the average growth law at the jamming transition as
\begin{equation}
    \frac{\textrm{d}R^3}{\textrm{d}t} \propto \ln\left(\frac{R}{\ell}\right)
\label{kissinglaw}
\end{equation}
plus a $\ln(\alpha)$ term with $\alpha=\mathcal{O}(1)$ that would depend on the bubble size distribution and can be dropped since $R\gg\ell$. Note that $\ell$ enters as a feature of the bubble packing microstructure at jamming not present in the very-wet dilute-bubble limit, where the average bubble size is the one and only relevant length scale.

The kissing-spheres average growth law, Eq.~(\ref{kissinglaw}), can be integrated by separating variables. This gives time versus average radius as
\begin{equation}
    t \approx \frac{1}{\mathcal{D}} \left[ \frac{R^3}{\ln(R/\ell)} - \frac{R_o^3}{\ln(R_o/\ell)} \right].
\label{kissing}
\end{equation}
for $R\gg\ell$. While this cannot be inverted for $R(t)$ versus $t$ exactly \cite{jacob}, it immediately reveals that the average radius grows only logarithmically faster than the $\beta=1/3$ power-law expectation for the very wet limit.  Indeed, as shown in Fig.~\ref{fig_kissing}, these two forms are indistinguishable in their comparison with the wettest foam data recently obtained aboard the ISS~\cite{CRAS2023} using the Soft Matter Dynamics apparatus~\cite{Born2021}. More precise data, taken over an impracticably long duration, would be needed to discriminate between Eq.~(\ref{kissing}) and an asymptotic 1/3 power law. Thus, Eq.~(\ref{kissing}) neatly explains the confounding discovery in Ref.~\cite{Isert2013} of $\beta \rightarrow 1/3$ near the jamming transition rather than in the dilute bubble limit.

%-------------------------------------------------
%-------------------------------------------------
\section{Coarsening above jamming}

\subsection{Truncated spheres growth law}

%=====================
\begin{figure}[t!]
\includegraphics[width=3.0in]{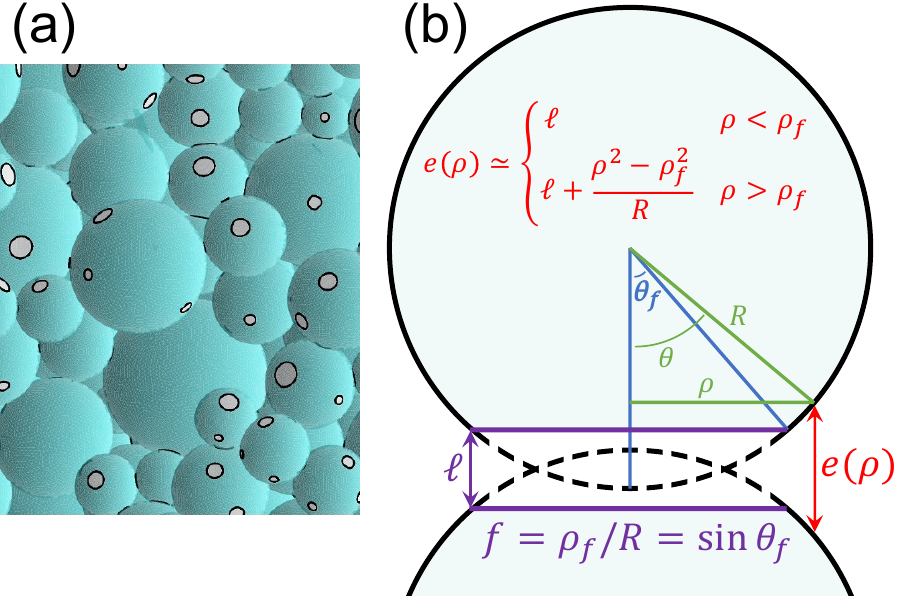}
\caption{Microstructure for wet foams that are jammed and polydisperse: (a) The bubbles are nearly spherical and their contacts are nearly circular soap films \cite{AndyImage}. (b) The average bubble radius $R$ is greater than the film radius $\rho_f$, which in turn is much greater than the film thickness $\ell$.  The ratio $f=\rho_f/R$ of film to bubble radii vanishes at the jamming transition, and the surface-surface separation $e(\rho)$ versus axial distance $\rho$ may be approximated by the displayed piecewise function for the region where most gas transport occurs.
}
\label{fig_bubbles}
\end{figure}
%=====================

The same border-crossing approach may now be extended to compute the diffuse gas current between bubbles compressed into contact above the jamming transition. In this regime, the bubbles are nearly spherical and their contacts are nearly-circular films of radius $\rho_f$ and thickness $\ell$. This geometry is illustrated in Fig.~\ref{fig_bubbles}a for a simulated polydisperse random packing (see Refs.~\cite{Hohler2019, Hutzler2020} for similar striking images for crystalline packings). As bubbles are progressively compressed with decreasing wetness, the film radius $\rho_f$ grows while $\ell$ and $R$ remain nearly constant. Thus, the distance to the jamming transition at $f=0$ is solely parameterized by the dimensionless ratio $f=\rho_f/R$. 

For average coarsening behavior in a self-similar scaling state, it's enough to consider gas transport between two bubbles of characteristic average radii $\approx R$, as per Mullins' argument and per going from Eqs.~(\ref{Is}-\ref{Dphi}) to Eq.~(\ref{kissinglaw}) above. Following the same border-crossing argument that gave Eq.~(\ref{Is}), the average volumetric growth is calculated as the integral of diffuse gas flux versus the area $2\pi\rho \textrm{d}\rho$ of the infinitesimal ring centered between the bubbles from axial distance $\rho$ to $\rho+\textrm{d}\rho$ through which it crosses.  This flux is proportional to the characteristic difference in dissolved gas concentration outside two neighboring bubbles, $\Delta\phi\propto 1/R$, controlled by Laplace's Law.  In the stratified approximation, the border-crossing flux at at axial distance $\rho$ is also inversely proportional to the separation $e(\rho)$ of the bubbles' surfaces. The average volumetric growth rate is thus
\begin{equation}
    \frac{d R^3}{\textrm{d}t} \propto
    \int_0^{\rho_f+aR} \left(\frac{\Delta\phi}{e(\rho)}\right)2\pi\rho \textrm{d}\rho.
\label{bordercrossingfacets}
\end{equation}
where $aR$ with $a=\mathcal{O}(1)$ imposes a cutoff beyond which there is negligible flux.

The true form of $e(\rho)$ in Eq.~(\ref{bordercrossingfacets}) depends on details of the actual bubble-bubble contact microstructure, which is different for different pairs of bubbles in a polydisperse packing. Far enough above jamming, it won't even be axially symmetric.  Perhaps the simplest reasonable approximation is to imagine the bubbles as truncated spheres, as depicted in Fig.~\ref{fig_bubbles}b.  The corresponding form of $e(\rho)$ may then be taken as $\ell$ for $\rho<\rho_f$ and $\ell+(\rho^2-\rho_f^2)/R$ for $\rho>\rho_f$. This gives an average volumetric growth rate of
\begin{eqnarray}
    \frac{\textrm{d}R^3}{\textrm{d}t} &\propto&  \ln\left[1+a(a+2f)\frac{R}{\ell}\right] + f^2\frac{R}{\ell} \label{circfacetsfull} \\
    &\approx& \ln\left(\frac{R}{\ell}\right) + f^2\frac{R}{\ell} \label{circfacets}
\end{eqnarray}
where the first expression is an exact evaluation of Eq.~(\ref{bordercrossingfacets}) and the approximation follows for $a(a+2f)=\mathcal{O}(1)$ and $R\gg\ell$. The log term comes from the $\rho>\rho_f$ border-crossing portion of the integral. The second term comes from the $\rho<\rho_f$ film portion of the integral. The latter vanishes at $f=0$, thus recovering the derivation of Eq.~(\ref{kissinglaw}) at the jamming transition.  Note that Eq.~(\ref{circfacets}) features three length scales that account for three different features of the bubble-packing microstructure of jammed foams: the average bubble radius $R$, the average film radius $\rho_f = fR$, and the film thickness $\ell$. Of these, only the film radius changes appreciably with liquid content -- through the value of $f$.  See Appendix-B for how Eq.~(\ref{circfacets}) changes for other bubble-bubble contact microstructure models of $e(\rho)$ in two and three dimensions.

\subsection{Bubble radius versus time}

\begin{figure}[t!]
\includegraphics[width=3.20in]{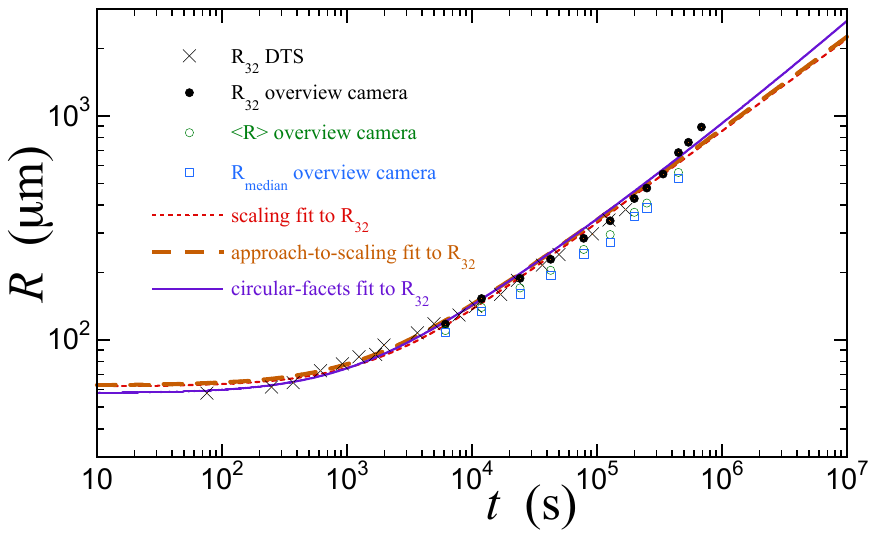}
\caption{Average bubble radius versus time data for a foam with $\varepsilon=40\%$ liquid.  The crosses and solid circles are for $R_{32}=\langle R^3\rangle/\langle R^2\rangle$, respectively obtained from diffuse transmission spectroscopy (DTS) and the overview camera, as shown Fig.~10 of Ref.~\cite{CRAS2023}. The open circles and squares are for the average and median radii, respectively, from the overview camera. The dotted and dashed curves represent fits of the combined $R_{32}$ data to Eq.~(\ref{approach}) with and without $\delta$ set to zero, respectively. The resulting exponent values are collected in Fig.~\ref{fig_exponents}. The solid curve represents numerical integration of Eq.~(\ref{circfacets}) using $\ell=30$~nm and $f=0.036$, per Fig.~\ref{fig_exponents}, starting from fixed initial conditions of $R_{32}(10~\text{s})=58~\mu$m; the proportionality constant was adjusted to $21~\mu\text{m}^3$/s to match the combined $R_{32}$ data.}
\label{fig_beta}
\end{figure}

While the ``truncated spheres" growth law of Eq.~(\ref{circfacets}) cannot be integrated analytically for time versus the average bubble radius, it may be integrated numerically and plotted as average radius versus time. An example is shown in Fig.~\ref{fig_beta} in comparison with a data set from Fig.~10 of Ref.~\cite{CRAS2023} for a $\varepsilon=40$\% wet foam measured aboard ISS. For this, $\ell=30$~nm and $f=0.036$ are taken (the latter from Fig.~\ref{fig_exponents}b), along with suitable initial conditions and proportionality constant in order to match $R_{32}(t)$ data obtained with DTS at early times and with the overview camera at late times (shown by $\times$ and $\bullet$ respectively). As seen, the resulting numerical integral of Eq.~(\ref{circfacets}) is certainly not a power law --as expected-- but matches the $R_{32}(t)$ data quite well.

%=====================
%=====================

Besides being rooted in a physical picture of microstructure in terms of the $\{\ell, \rho_f, R\}$ length scales, the new growth law actually provides a slightly better description of the data than fits to Eq.~(\ref{approach}). Fixing $\delta=0$ or allowing it to adjust give respective effective exponents of $\beta=0.41$ and 0.42, clearly intermediate between the wet and dry limits. As seen in Fig.~\ref{fig_beta}, the combined $R_{32}(t)$ data don't asymptote well to a power law and the resulting Eq.~(\ref{approach}) fits don't have as much upward curvature on a log-log plot as the data. This subtle feature is captured by the numerical integration of Eq.~(\ref{circfacets}). In an alternative analysis~\cite{PasquetSM2023}, the final behavior of our same $\varepsilon=40\%$ dataset from Ref.~\cite{CRAS2023} was reported to be both $\mathcal{D}t^{0.36}$ and consistent with $\beta=1/3$ for the purpose of analyzing the overall rate constant $\mathcal{D}$.

As an aside, the overall rate constant in growth laws such as Eqs.~(\ref{approach},\ref{circfacets}) must vary with materials parameters of surface tension, gas diffusivity, Henry's Law constant, and film thickness -- all as set by specific choice of foaming system. Less obviously, the proportionality constant must also depend on details of the bubble size and shape distributions as set by liquid content and the universal nature of the statistical self-similar scaling state -- presuming it exists. This is an unsolved problem (see Refs.~\cite{Sabik2024, Morgan2026} for relevant recent state of art), but there are two solid touchstones. The first comes from the field of jamming, which established universal behavior for the coordination number $z$ near the transition for repulsive spheres in D~dimensions. Namely, independent of the form of the particle interaction or size distribution, the coordination number varies dramatically near jamming as $z-2\textrm{D}\propto\sqrt{\phi-\phi_c}$ \cite{BubblePRL, BubblePRE, Epitome, Wyart2005}. The second touchstone comes from the assumption of self-similarity and the von~Neumann Law $dA_n/dt=K_o(n-6)$ for the rate of area change for $n$-sided bubbles in an ideally dry two-dimensional foam~\cite{VonNeumann, Mullins56}. Combining self-similarity and von~Neumann gives an exact identity for the growth rate of the average bubble area:
\begin{equation}
    \frac{d\langle A\rangle}{\textrm{d}t} = 2K_o\frac{\langle A\rangle^2}{\langle A^2\rangle} \bigg(\langle\langle n\rangle\rangle-6 \bigg)
\label{eq_avgvonNeumann}
\end{equation}
where $\langle\langle n\rangle\rangle \approx 6.7$ is the area-weighted average side number in the scaling state \cite{Stavans93, RothPRE2013, ChiecoFSM22}, distinct from the number-weighted average side number $\bar{n}=6$ \cite{WeaireHutzlerBook, Cantat2013, Langevin2020}; see Ref.~\cite{ChiecoFSM22} for related identities. Though the form of the bubble area distribution is unknown, it enters via $\langle A\rangle^2/\langle A^2\rangle$. Size and topology distributions enter together via the value of $\langle\langle n\rangle\rangle$ in the scaling state. Any theory for $\mathcal{D}$ in Eq.~(\ref{approach}) must somehow similarly account for its dependence on the bubble size and shape/topology distributions in the assumed scaling state, and how they vary with wetness. This is an outstanding challenge not considered in Ref.~\cite{PasquetSM2023}, but \hbox{--fortunately--} is irrelevant to the exponent analyses below.

%-------------------------------------------------
\subsection{Coarsening exponent versus film radius}

While Eq.~(\ref{circfacets}) predicts non-power law average growth, it can nevertheless provide insight into an effective approximate coarsening exponent, $\beta$, for comparison with empirical power-law fits. For this, note that the two terms in Eq.~(\ref{circfacets}) respectively give exponents of roughly $1/3$ and exactly $1/2$ if they were to act alone.  Therefore, an effective value of $\beta$ could be estimated as the average of these values using the respective terms as weights.  A better-justified value is given by self-consistently requiring the right hand side of $\beta = \textrm{d}\ln R/\textrm{d}\ln t$ for the given growth law to be independent of time to first order (see Appendix-A for calculation details). The resulting effective exponent is the reciprocal of a weighted average of the reciprocal of the individual exponents:
\begin{equation}
    \beta = \frac{  \ln\left(\frac{\Bar{R}}{\ell}\right)  + f^2\frac{\Bar{R}}{\ell} }{  \left[3\ln\left(\frac{\Bar{R}}{\ell}\right) -1\right] + 2f^2\frac{\Bar{R}}{\ell} }
\label{beta}
\end{equation}
where $-1$ in the denominator is consistent with the log term not giving an exact 1/3 power law contribution.  Here, $\Bar{R}$ is the time-average of the sample-average bubble size, with the former taken over the duration of the observations.  In typical experiments $R$ grows by only a decade or so; therefore, one might take $\Bar{R}$ as the geometric average of the characteristic radii at the earliest and latest times. The expectation is that longer-duration growth data would appear less power law-like, in accord with Eq.~(\ref{circfacets}), and that results for $\beta$ would depend on fitting range. 

\begin{figure}[t!]
\includegraphics[width=3.20in]{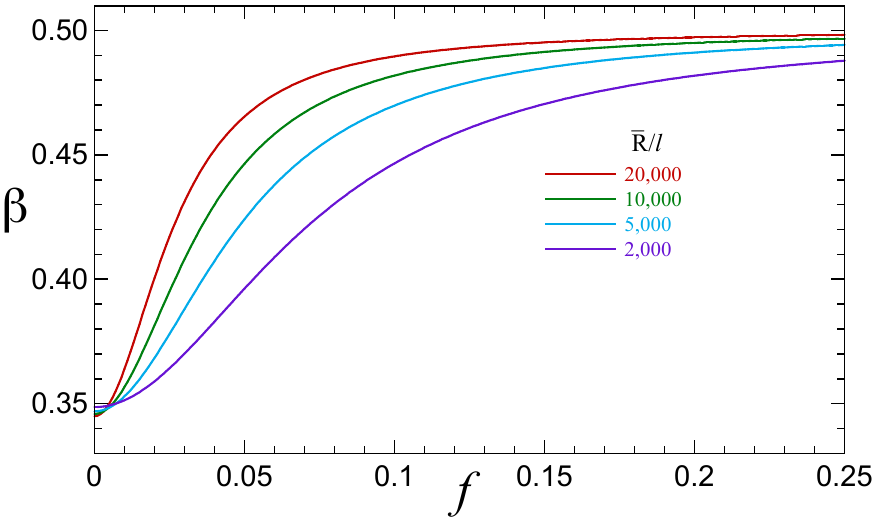}
\caption{Effective coarsening exponent $\beta$ versus the ratio $f$ of film to bubble radii, according to Eq.~(\ref{beta}).  The prediction is shown for several values of the time-average of the sample-average bubble radius $\bar R$ divided by film thickness $\ell$. The Ref.~\cite{Isert2013, CRAS2023} experiments correspond to $\bar R/\ell = 6700$ and 13000, respectively.}
\label{fig_betaf}
\end{figure}

For a given foam system with known $\bar R/\ell$, empirical fitting results for the effective exponent can now be understood physically solely in terms of the dimensionless parameter $f$ specifying the average film radius $\rho_f=fR$.  This is illustrated in Fig.~\ref{fig_betaf}, where the Eq.~(\ref{beta}) prediction for $\beta$ versus $f$ is plotted for a few values of $\bar R/\ell$. The bigger $\bar R/ \ell$, the more abruptly $\beta$ is predicted to fall towards $1/3$ on approach to the jamming transition at $f=0$. Note that at jamming, for nearly-kissing spheres, the effective exponent is $\beta = \ln(\bar R/\ell)/[3\ln(\bar R/\ell)-1]$, which is larger than $1/3$ by an amount that decreases with increasing $\bar R/\ell$.  Going above jamming, a reasonable upper limit for $f$ is about $f_{max}=0.5$, based on $4\pi R^2 \approx z \pi (fR)^2$ and a coordination number of $z\approx 14$ as appropriate for ideally dry random foams where bubble surfaces consist entirely of films \cite{Matzke, HilgenfeldtKraynikKoehlerStone01, Monnereau2001}. With increasing $f$, the predicted $\beta$ is thus seen in Fig.~\ref{fig_betaf} to reach 1/2 well below $f_{max}$.  These predicted trends are in qualitative agreement with experiments~\cite{Isert2013, CRAS2023}, including the supposition in Ref.~\cite{PasquetSM2023} of a step-function crossover at adhesive jamming.

%=====================
%=====================

\subsection{Coarsening exponent versus liquid fraction}

The prediction of Eq.~(\ref{beta}) can be further compared with experimental exponent data by connecting $f$ with liquid fraction. This will be done for two foam systems, one where drainage was eliminated by magnetic levitation \cite{Isert2013} and one where drainage was eliminated by microgravity aboard the ISS~\cite{Born2021, CRAS2023}. The bubbles in the levitated system have $\bar R\approx 200$~$\mu$m and are relatively nonadhesive. The film thickness is roughly $\ell\approx 15$~nm \cite{Bergeron1992}, giving $\bar R/\ell\approx 13000$. Since exponent results were not tabulated, the values of $\beta$ versus $\epsilon$ were extracted directly from Fig.~4 of Ref.~\cite{Isert2013} using WebPlotDigitizer~\cite{webplotdigitizer}. The bubbles in the microgravity system also have $\bar R \approx 200~\mu$m but are noticeably adhesive and have thicker films of about $\ell \approx 30$~nm and hence $\bar R/\ell\approx 6700$ \cite{Bergeron1992, Galvani23, PasquetSM2023}.  Slightly different film thicknesses were reported in Ref.~\cite{SchulzeSchlarmann2006}. 

Exponent results for the microgravity system are deduced as follows.  Since the growth predicted by Eq.~(\ref{circfacets}) is not a pure power law, and also since the microgravity foams have surprising transients that decay as the foam coarsens~\cite{Galvani23}, caution is needed. To obtain a conservatively inclusive range of effective exponents, fits of Eq.~(\ref{approach}) with and without fixing $\delta$ to zero are made to the combined $R_{32}(t)$ data published in Fig.~10 of Ref.~\cite{CRAS2023}.  This was exemplified in Fig.~\ref{fig_beta}. To further test for self-similarity, the mean and median sphere-equivalent radii are additionally computed from the overview camera data.  This was also exemplified in Fig.~\ref{fig_beta}, where it appears that the various radii may become proportionate at late times -- consistent with the decay of transients on approach to self-similarity.  The mean and median radii data are fit to Eq.~(\ref{approach}) with variable $\delta$.  Thus, four different effective exponent estimates are obtained at each liquid fraction for the microgravity system. In an alternative analysis, exponents from fits of $R(t)=\mathcal{D}t^\beta$ to the presumed asymptotic final factor of 2--4 growth of $R_{32}(t)$ are tabulated in Ref.~\cite{PasquetSM2023}.

\begin{figure}[t!]
\includegraphics[width=3.20in]{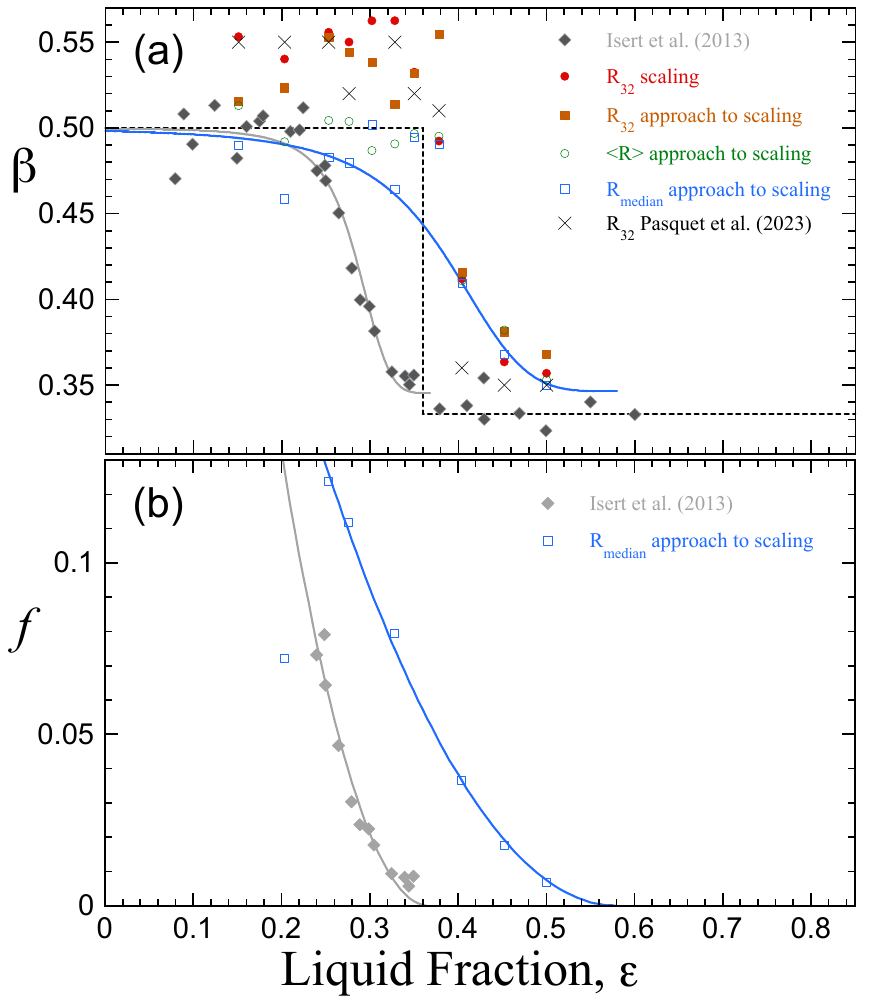}
\caption{(a) Effective coarsening exponents versus liquid fraction. The diamonds are from Fig.~4 of Isert \textit{et al.}~\cite{Isert2013}.  All other symbols are exponents for growth data partially published in Ref.~\cite{CRAS2023}: solid circles and squares for fits of Eqs.~(\ref{approach}) to $R_{32}$; open circles and squares are from fits to average and median radii from the overview camera; $\times$ symbols are from Table~1 of Pasquet \textit{et al.}~\cite{PasquetSM2023}.  The dotted line segments are the step function $1/2$ for $\varepsilon<0.36$ and $1/3$ for $\varepsilon>0.36$. (b) Ratio $f=\rho_f/R$ of average film to bubble radii as deduced from the exponent data using Eq.~(\ref{beta}).  The solid curves are fits to parabolic forms given in the text, and are used to generate the solid curves in (a).
}
\label{fig_exponents}
\end{figure}

All the various exponents are plotted versus liquid fraction $\varepsilon$ in Fig.~\ref{fig_exponents}a.  With increasing wetness, $\beta$ values for the levitated system decrease from around 1/2 to around 1/3 near random close packing per the now-resolved motivating puzzle~\cite{Isert2013}.  For the microgravity system, $\beta$ values are considerably more scattered, crossing from around 1/2 to slightly above 1/3 at a higher liquid fraction. This could be due to bubble-bubble adhesion \cite{Galvani23, PasquetSM2023}, which could pull bubbles into closer contact and perhaps cause the ratio of film to bubble radii to be a nonzero constant, $f_0$, even in the wet limit. Another possibility is that $1/3$ isn't reached because the nearly-kissing sphere growth law is logarithmically faster than a $1/3$ power law, per the $-1$ term in Eq.~(\ref{beta}). 

%=====================
%=====================

Values of $f$ may now be deduced from the collected exponent results, using Eq.~(\ref{beta}) and the above values of $\bar R/\ell$.  This is shown in Fig.~\ref{fig_exponents}b for both foaming systems, where for the microgravity system $f$ is shown only from $\beta$ for fits to the median bubble size versus time, since it has the least scatter and remains mostly below 1/2.  As anticipated from Fig.~\ref{fig_betaf}, the $f$ values are well below $f_{max}$ as $\beta$ begins to fall below 1/2 with increasing $\varepsilon$. For even wetter samples, $f$ appears to decrease toward zero with upward curvature on approach to unjamming for both data sets -- but at different critical liquid fractions $\varepsilon_c$.  The simplest empirical form is thus $f(\varepsilon) = f_2(\varepsilon_c-\varepsilon)^2$ below $\varepsilon_c$ and $f(\varepsilon)=0$ above. Fits give \{$f_2=4.8\pm0.8, \varepsilon_c=0.37\pm0.01$\} for the levitated system and \{$f_2=1.2\pm0.1, \varepsilon_c=0.58\pm0.01$\} for the the microgravity system.  These forms correspond to the solid curves in Fig.~\ref{fig_beta}a, which match the respective $\beta$ versus $\varepsilon$ data. Thus, the observed variation of the exponents is directly understood though variation of the microstructure in terms of how the film radius changes with wetness.

%----------------------------------------------
\section{Conclusion}

In conclusion, the success of Eqs.~(\ref{kissing}-\ref{beta}) based on border-crossing physics fills a void in the literature by providing physical insight and a general theoretical framework for understanding diffusive growth in terms of microstructure and how it changes between the dry and jamming limits.  The effective exponent $\beta$ crosses from 1/2 to roughly 1/3 for increasingly wetter foams as film radii shrink and border crossing becomes more important.  The findings reconcile different observed forms for $\beta$ vs liquid content in terms of system-specific details such as film thickness, degree of adhesion, or location of the jamming transition, and are applicable to a broad range of phase-separating materials with analogous microstructure \cite{Mullins1986, Stavans93}.  One lesson is that border-crossing physics gives rise to non power-law average growth laws, even if data is empirically well-described by an effective exponent less than 1/2.

This work also opens myriad lines of further research.  Besides highlighting the need for even more and better growth data and effective exponents near the jamming transition, it would be interesting if the results seen for $f(\varepsilon)$ could be justified by combining ideas from the field of jamming with knowledge of how foam microstructure arises from surface tension and interfacial forces. This is nontrivial because of the diversity of bubble sizes and shapes as well as coordination number, and how they all vary with distance to jamming. The empirical connection of liquid fraction to the average fraction of a bubble covered with films could be helpful~\cite{Hohler2021}, but needs to be supplemented with knowledge of the coordination number.  Other extensions would be to include a contact angle between adhering bubbles and the slight variation of $\ell$ with liquid content. The same border-crossing approach can also be invoked to predict growth rates and effective exponents near the dry limit and for two-dimensional foams (see Appendix-B). All these predictions could be tested and refined by incorporating border-crossing expressions for diffusive gas flux into simulations of accurate bubble-scale microstructure for wet foams, \textit{e.g.}~by extension of Refs.~\cite{GlazierAndersonGrest90, Brakke, Lazar2011, Graner2012, Graner2015, Peskin2014, Hohler2019, Hutzler2020, Sabik2024, Morgan2026}.

%----------------------------------------------
%----------------------------------------------

\begin{acknowledgments}
I thank Fran\c{c}ois Graner, Reinhard H\"ohler, Dominique Langevin, Jacob Morgan, and Arnaud Saint-Jalmes for helpful discussions; Sylvie Cohen-Addad and Olivier Pitois for assistance with ISS data from our collaborative Foam-C project \cite{Born2021, Galvani23, CRAS2023}; Andy Kraynik for permission to use Fig.~\ref{fig_bubbles}a \cite{AndyImage}; and NASA for supporting my work on foam behavior versus liquid content since 1992, including most recently under grant number 80NSSC21K0898. In addition I thank the Center for Computational Biology at the Flatiron Institute, a division of the Simons Foundation, for support and hospitality while a portion of this research was performed. This research was also supported in part by grant NSF PHY-2309135 to the Kavli Institute for Theoretical Physics (KITP).
\end{acknowledgments}

%----------------------------------------------
%----------------------------------------------

\bibliography{FoamRefs}

%----------------------------------------------
%----------------------------------------------

\clearpage

\appendix

\section{Calculation of Effective Exponent for Non Power-Law Average Growth}\label{AppBeta}

To cover all the cases in main text and below, let's consider an average coarsening growth law of form
\begin{equation}
    \frac{d R^3}{\textrm{d}t} = a_0\ln\left(\frac{R}{\ell}\right) + a_1\sqrt{\frac{R}{\ell}} + a_2 \frac{R}{\ell}.
    \label{gengrowth}
\end{equation}
Since this does not give power law growth, the slope
\begin{eqnarray}
    \beta(t) &=& \frac{\textrm{d}\ln R}{\textrm{d}\ln t} = \frac{t}{R}\frac{\textrm{d}R}{\textrm{d}t} \\
    &=& \frac{t}{3R^3}\left[ a_0\ln\left(\frac{R}{\ell}\right) + a_1\sqrt{\frac{R}{\ell}} + a_2 \frac{R}{\ell}\right] \label{betat}
\end{eqnarray}
of $R(t)$ vs $t$ on a log-log plot is not a constant. But for some degree of self-consistency, the value of $\beta$ may be taken such that the time derivative of the right-hand side of this equation is independent of time to first order in~$t$.  This is achieved by computing $\textrm{d}\beta(t)/\textrm{d}t$ from Eq.~(\ref{betat}) in terms of \{$R$, $\textrm{d}R/\textrm{d}t$, $t$\}, substituting in $\textrm{d}R/\textrm{d}t$ from Eq.~(\ref{gengrowth}) and $t = 3\beta R^3 / ( a_0\ln R + a_1\sqrt{R} + a_2 R)$ from Eq.~(\ref{betat}), equating the resulting expression of $\textrm{d}\beta/\textrm{d}t$ to zero, and then solving for $\beta$.  This gives an effective exponent in terms of $R$ as
\begin{equation}
    \beta = \frac{a_0 \ln\left(\frac{R}{\ell}\right) + a_1 \sqrt{\frac{R}{\ell}} + a_2 \frac{R}{\ell}}{a_0\left[3\ln\left(\frac{R}{\ell}\right)-1\right] + \frac{5}{2}a_1\sqrt{\frac{R}{\ell}} + 2a_2\frac{R}{\ell}}
    \label{beta012}
\end{equation}
Thus, for an average growth law that is a sum of terms that would individually give exponents $\beta_i$ if acting alone, the effective exponent is given as $1/\beta$ equal to the weighted average of $1/\beta_i$ using their respective terms in the growth law as weights. Nontrivially, the $-1$ in Eq.~(\ref{beta012}) arises from the $\ln R$ term in Eq.~(\ref{gengrowth}) because it is associated with growth that is logarithmically faster than a $1/3$ power law.

\section{Other Border-Crossing Predictions}

The border-crossing method is applied below to compute average growth laws and the consequent variation of effective exponents with wetness for several other situations for both 2D and 3D foams.

\subsection{Alternative pictures of bubble-bubble contact microstructure near jamming in 3D}

\subsubsection{Zero contact angle}

\begin{figure}[t!]
\includegraphics[width=3.0in]{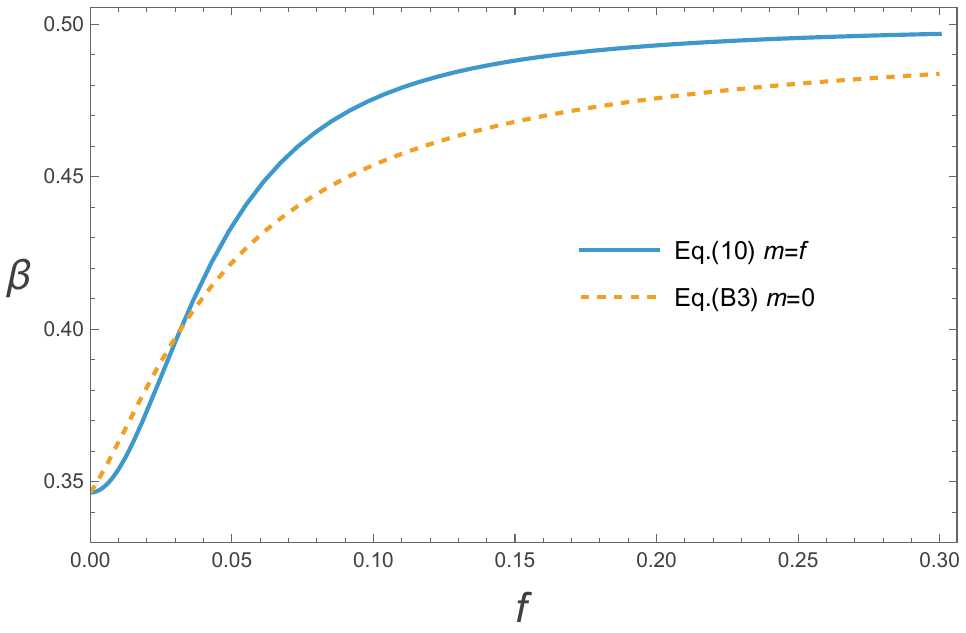}
\caption{Effective exponent versus the ratio $f=\rho_f/R$ of film to bubble radius as predicted by the two different growth laws.  These are shown for $\bar R/\ell=6700$, corresponding to the microgravity experiments.}
\label{figa_beta}
\end{figure}

The complex microstructure of bubble-bubble contacts in a disordered 3D wet foam can only be found by simulation. The sketch in Fig.~\ref{fig_bubbles}b is a simplified caricature, treating the bubbles as truncated spheres. In this picture, the film flares out with a nonzero contact angle $\theta_f$ and slope $m=\tan\theta_f$ given by $m=2\textrm{d}e/\textrm{d}\rho|_{\rho_f^+}=f$.  For the spherical and film portions of the surface to meet more smoothly, with zero contact angle, the separation of bubble-bubble surfaces at axial distance $\rho$ could instead be taken by $e(\rho)=\ell$ for $\rho<\rho_f$ and by $e(\rho)=\ell+(\rho-\rho_f)^2/R$ for $\rho>\rho_f$. This gives an average volumetric growth rate of
\begin{eqnarray}
    \frac{\textrm{d}R^3}{\textrm{d}t} &\propto& \ln\left(1+a^2\frac{R}{\ell}\right) + \nonumber \\
    & & 2\sqrt{\frac{f^2R}{\ell}}\arctan\left(a\sqrt{\frac{R}{\ell}}\right) 
    + \frac{f^2R}{\ell}    
    \label{circfacetsV2full} \\
    &\approx& \ln\left(\frac{R}{\ell}\right) + \pi\sqrt{\frac{f^2R}{\ell}} + \frac{f^2R}{\ell}
    \label{circfacetsV2}
\end{eqnarray}
where the first expression is an exact evaluation of Eq.~(\ref{bordercrossingfacets}) and the approximation holds for $a=\mathcal{O}(1)$ and $R\gg\ell$.  This growth law is similar to Eq.~(\ref{circfacets}), differing only by a third term $\propto\sqrt{R/\ell}$ that, if it acted alone, would give a growth exponent of 2/5. According to Eq.~(\ref{betat}), the effective exponent is thus
\begin{equation}
    \beta = \frac{  \ln\left(\frac{\Bar{R}}{\ell}\right)  + \pi\sqrt{\frac{f^2R}{\ell}} + f^2\frac{\Bar{R}}{\ell} }{  \left[3\ln\left(\frac{\Bar{R}}{\ell}\right) -1\right] + \frac{5}{2}\pi\sqrt{\frac{f^2R}{\ell}} + 2f^2\frac{\Bar{R}}{\ell} }
\label{betaV2}
\end{equation}
This is compared with Eq.~(\ref{beta}) in Fig.~\ref{figa_beta} for the $\bar R/\ell = 6700$ corresponding to the ISS microgravity experiments.  Under these conditions, the two different pictures of bubble-bubble contact microstructure give similar effective exponents in terms of of $f$ near the jamming transition.  But they differ noticeably above the transition where $\beta$ is closer to 1/2 than to 1/3.

%=====================
%=====================

\subsubsection{Variable contact angle}

To account for a nonzero contact angle $\theta$ due to adhesion, the separation of bubble-bubble surfaces at axial distance $\rho$ could be approximated as $e(\rho)=\ell$ for $\rho<\rho_f$ and $e(\rho)=\ell+2m(\rho-\rho_f)+(\rho-\rho_f)^2/R$ to leading order for $\rho>\rho_f$. Here $m=\tan\theta$ is the slope of the bubble surfaces with respect to the plane of the film, at the contact line. This form is still an approximation -- the actual form of $e(\rho)$ is set by minimizing the total energy as comprised of liquid-gas and film tensions as well as contributions from the crossover region where the film thickness increases from equilibrium to beyond the range of the effective interface potential (which has an attractive piece, making the contact angle nonzero, and a repulsive piece, preventing film rupture). By contrast with two otherwise-isolated bubbles in equilibrated contact underwater, in a foam the film radius $\rho_f=fR$ depends on the liquid content. The border-crossing average growth law for the simplified form of $e(\rho)$ is
\begin{eqnarray}
    \frac{\textrm{d}R^3}{\textrm{d}t} &\propto& \ln\left[1+a(a+2m)\frac{R}{\ell}\right] +2\frac{f-m}{M_T}\times \nonumber \\
    &&\left(\arctan\frac{a+m}{M_T}-\arctan\frac{m}{M_T}\right) + \frac{f^2R}{\ell} \label{circfacetsV3} \\ 
    &=& \ln\left[1+a(a+2m)\frac{R}{\ell}\right] + \frac{f-m}{M_L}\times \nonumber \\
    &&\ln\left(\frac{\frac{\ell}{R}+am+aM_L}{\frac{\ell}{R}+am-aM_L}\right) + \frac{f^2R}{\ell} \label{circfacetsV4}
\end{eqnarray}
where $M_T=\sqrt{(\ell/R)-m^2}$ and $M_L=\sqrt{m^2-(\ell/R)}$. These are different but equal expressions for the exact evaluation of Eq.~(\ref{bordercrossingfacets}) for the assumed $e(\rho)$. Both are real, but one more obviously than the other depending on the size of $m^2$ relative to $\ell/R$.  The first clearly reduces to Eq.~(\ref{circfacetsV2}) for $m=0$. Less obviously, they both reduce to Eq.~(\ref{circfacets}) for $m=f$ as required.
For $a=\mathcal{O}(1)$ and $R\gg\ell$, they expand to
\begin{eqnarray}
    \frac{\textrm{d}R^3}{\textrm{d}t} &\approx& \ln\left(\frac{R}{\ell}\right) + \pi\sqrt{\frac{f^2R}{\ell}} + \frac{f(f-2m)R}{\ell}
    \label{circfacetsV3a} \\
    \frac{\textrm{d}R^3}{\textrm{d}t} &\approx& \ln\left(\frac{R}{\ell}\right) + \frac{f-m}{m}\ln\left[\frac{4am^2R}{(a+2m)\ell}\right] \nonumber \\ && + \frac{f^2R}{\ell}
    \label{circfacetsV4a}
\end{eqnarray}
for $m^2\ll \ell/R$ and $m^2\gg\ell/R$, respectively. Note that the first approximation is identical to Eq.~(\ref{circfacetsV2}) except for the $-2f\theta$ term, which causes slower growth as expected because $e(\rho)$ grows faster for $\rho>\rho_f$ for nonzero contact angles (presuming the same $\ell$ -- though of course films are thinner when there is adhesion).  Note too that the cutoff $a$ drops out of the second approximation for $m\ll a$.

\begin{figure}[t!]
\includegraphics[width=3.0in]{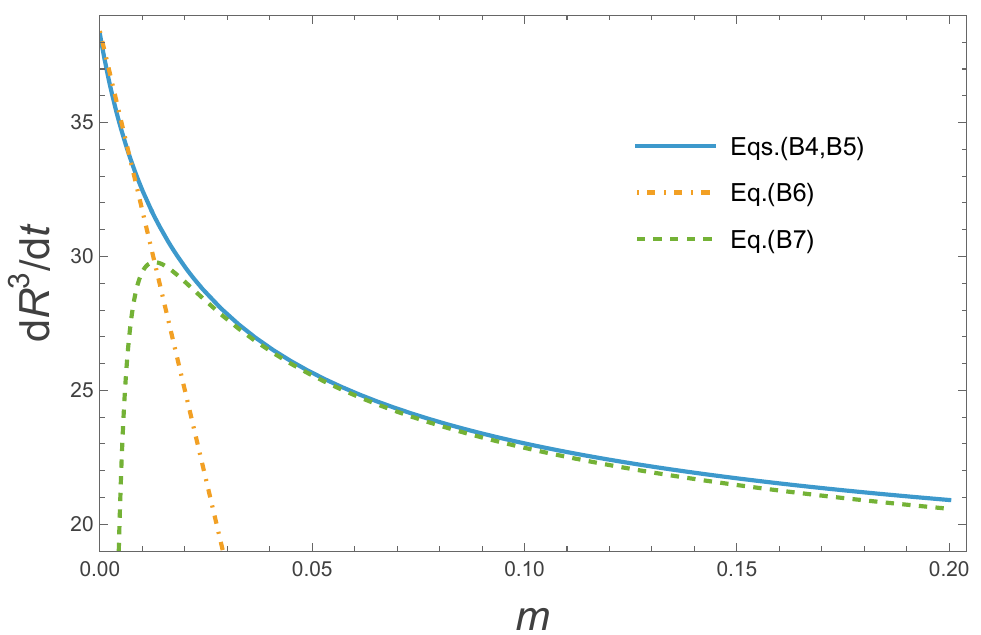}
\caption{Average volumetric growth rate vs tangent $m$ of contact angle, as given by the right-hand size of the specified equations.  These are shown for $R/\ell=6700$, $a=1$, and $f=0.05$.  The two approximations respectively hold for $m$ smaller and larger than $\sqrt{\ell/R}$, which equals 0.012 here.}
\label{figa_rate}
\end{figure}

The average volumetric growth rate as predicted by the right hand sides of Eqs.~(\ref{circfacetsV3},\ref{circfacetsV4}) is plotted versus the tangent $m$ of the contact angle in Fig.~\ref{figa_rate} for $R/\ell=6700$, $a=1$, $f=0.05$. As the contact angle increases from 0 to $11^\circ$, $m$ increases from 0 to 0.2 and the predicted growth rate decreases by a factor two -- first quickly and then more gradually. For $m$ smaller and larger than $\sqrt{\ell/R}=0.012$, this behavior is respectively captured by the Eq.~(\ref{circfacetsV3a},\ref{circfacetsV4a}) approximations.

%=====================
%=====================

\subsubsection{General form for growth law?}

In all three models considered for $e(\rho)$, the resulting large-$R$ average volumetric growth laws feature $\ln(R/\ell)$ and $f^2 R/\ell$ terms from contributions due to border crossing and the films, respectively. They differ by the presence or absence of additional terms depending on details of how the gas-liquid interfaces flare out away from their junction with the film.  So it's natural to hypothesize that, if the actual bubble-bubble contact microstructures were known and fluxes were averaged over a scaling distribution, the average volumetric growth law --including its dependence on $f$-- would be of good approximate form
\begin{equation}
    \frac{\textrm{d}R^3}{\textrm{d}t} \propto \ln\left(\frac{R}{\ell}\right) + \kappa_1\sqrt{\frac{f^2R}{\ell}} + \kappa_2\frac{f^2R}{\ell}
\label{circfacetsV5}
\end{equation}
where the dimensionless constants $\{\kappa_1,\kappa_2\}$ depend on specific structural details of the bubble-bubble contacts, including the contact angle between adhering bubbles as well as the bubble size and shape distributions in the scaling state (\textit{c.f.} Eq.~(\ref{eq_avgvonNeumann})).  This gives non-power law growth, of course.  But per Appendix~\ref{AppBeta}, the corresponding effective exponent would be
\begin{equation}
    \beta = \frac{  \ln\left(\frac{\Bar{R}}{\ell}\right) + \kappa_1\sqrt{\frac{f^2\Bar{R}}{\ell}} + \kappa_2\frac{f^2\Bar{R}}{\ell} }{  \left[ 3\ln\left(\frac{\Bar{R}}{\ell}\right)-1\right] + \frac{5}{2}\kappa_1\sqrt{\frac{f^2\Bar{R}}{\ell}} + 2 \kappa_2\frac{f^2\Bar{R}}{\ell}  }
\label{betaV3f}
\end{equation}
as set by the weighted average of the three contributions. The challenge for future work would be to find both the connection of $f$ to $\varepsilon$ as well as the values of $\kappa_i$ -- all in terms of bubble-bubble contact microstructure distributions. In absence of such guidance, Eq.~(\ref{circfacets}) is used in the main text since it's simpler than Eqs.~(\ref{circfacetsV2},\ref{circfacetsV5}).

% https://arxiv.org/abs/2304.00415v1

\subsection{Near the dry limit in 3D}

For slightly wet foams, the bubbles are nearly polyhedral and the Plateau borders inflate with liquid, having length $\sim R$ and radius of curvature $r\sim \sqrt{\varepsilon} R$.  The latter follows from separation of length scales $\ell \ll r \ll R$ and the order of magnitude estimate for the liquid fraction as the sum of film, border, and vertex contributions: $\varepsilon \sim (R^2\ell + r^2 R + r^3)/R^3 \sim (r/R)^2$, ignoring numerical factors. The average volumetric growth rate is given by the sum of diffuse currents across films and borders as
\begin{eqnarray}
    \frac{\textrm{d}R^3}{\textrm{d}t} &\sim& \frac{D\Delta\phi}{\ell}R^2 + \frac{D\Delta\phi}{\langle e\rangle}rR \\
    &\sim& \frac{R}{\ell} + \kappa_1 \sqrt{ \frac{\sqrt{\varepsilon} R}{\ell} }
\end{eqnarray}
using $\Delta\phi\sim 1/R$, an average diffuse border thickness of $\langle e\rangle \sim \sqrt{\ell r}$ \cite{Schimming2017}, and inserting a dimensionless constant $\kappa_1$ that depends on yet-unknown details of the microstucture per above. This argument is equivalent to integrating the diffuse flux $D\Delta\phi/e(\rho)$ versus $2\pi\rho \textrm{d}\rho$ where the surface-surface separation $e(\rho)$ is $\ell$ for $\rho<\rho_f$ and $\ell+(\rho-\rho_f)^2/r$ for $\rho_f<\rho<\rho_f+ar$ where the film radius $\rho_f $ is of order $R$. The corresponding effective exponent follows as
\begin{equation}
    \beta = \frac{  \left(\frac{\Bar{R}}{\ell}\right) + \kappa_1\sqrt{\frac{\sqrt{\varepsilon}\Bar{R}}{\ell}}  }{  2\left(\frac{\Bar{R}}{\ell}\right) + \frac{5}{2}\kappa_1\sqrt{\frac{\sqrt{\varepsilon}\Bar{R}}{\ell}}  }
\label{betadry}
\end{equation}
per Appendix~\ref{AppBeta}.

\subsection{Near the dry limit in 2D}

Border-crossing expectations for the average coarsening behavior of two-dimensional foams may be developed using the same arguments as above. For the slightly wet case, the bubbles are nearly polygonal and the vertices, at which three films of length $\sim R$ meet, inflate with liquid to a radius of curvature given by $r\sim \sqrt{\varepsilon}R$. This follows from the liquid fraction estimate of $\varepsilon \sim (\ell R + r^2)/R^2 \sim (r/R)^2$. The average area grows as the sum of diffuse currents across films and vertices:
\begin{eqnarray}
    \frac{\textrm{d}R^2}{\textrm{d}t} &\sim& \frac{D\Delta\phi}{\ell}R + \frac{D\Delta\phi}{\langle e\rangle}r \\
    &\sim& \frac{1}{\ell} + \kappa_1 \sqrt{ \frac{\sqrt{\varepsilon} }{\ell R} }
\end{eqnarray}
where $\langle e\rangle \sim \sqrt{\ell r}$ is the effective diffuse thickness of the vertices \cite{Schimming2017} and $\kappa_1$ is a dimensionless constant that depends on details of the microstructure in the scaling state. The average radius thus grows with the same form as in 3D, with corresponding effective exponent given by Eq.~(\ref{betadry}).

\subsection{Near jamming in 2D}
For wet 2D foams of slightly jamming bubbles, the average area grows according to an integral like Eq.~(\ref{bordercrossingfacets}) but over $\textrm{d}\rho$ rather than $2\pi \rho \textrm{d}\rho$:
\begin{equation}
    \frac{\textrm{d}R^2}{\textrm{d}t} \propto
    \int_0^{\rho_f+aR} \left[\frac{D\Delta\phi}{e(\rho)}\right]\textrm{d}\rho.
\label{bordercrossingfacets2D}
\end{equation}
where $\rho_f = f R$ is film length, $\Delta \phi \sim 1/R$, and the bubble-bubble separation $e(\rho)$ at axial distance $\rho$ is $\ell$ for $\rho<\rho_f$ and either $\ell + (\rho^2-\rho_f^2)/R$ or $\ell + (\rho-\rho_f)^2/R$ for $\rho>\rho_f$. Both choices for the microstructure give
\begin{equation}
    \frac{\textrm{d}R^2}{\textrm{d}t} \sim \sqrt{\frac{1}{R\ell}} + \kappa_1 \frac{f}{\ell} 
\end{equation}
where $\kappa_1$ is again a dimensionless constant that depends on average microstructure in the scaling state.  The corresponding effective exponent is
\begin{equation}
    \beta = \frac{  \sqrt{\frac{\bar R}{\ell}} + \kappa_1 \frac{f\bar R}{\ell} }{\frac{5}{2}\sqrt{\frac{\bar R}{\ell}} + 2\kappa_1 \frac{f\bar R}{\ell} }
\label{betadry2D}
\end{equation}
By contrast with the 3D case, the coarsening exponent is 2/5 at jamming $(f=0)$. The crossover to $1/3$ would then happen only on approach to the very-wet dilute-bubble limit per ordinary expectation.

%----------------------------------------------
%----------------------------------------------

%\bibliography{FoamRefs}

\end{document}